\documentclass[12pt]{article}

\usepackage{a4}
\usepackage{amsfonts,amssymb,amsmath}
\usepackage{graphicx}
\usepackage{cite}
\newtheorem{theorem}{Theorem}

\newtheorem{lemma}[theorem]{Lemma}

\newenvironment{proof}[1][Proof]{\textbf{#1.} }{\ \rule{0.5em}{0.5em}}

\begin{document}
\title{Star products made (somewhat) easier}
\author{{V.G. Kupriyanov\thanks{E.mail: vladislav.kupriyanov@gmail.com} and 
D.V. Vassilevich\thanks{Also at Physics Department, 
St.Petersburg University, Russia. E.mail:
dvassil@gmail.com}} \\
Instituto de F\'{\i}sica, Universidade de S\~{a}o Paulo,\\ 
Caixa Postal 66318, CEP 05315-970, S.P. Brazil}
\date{\today                                        }
\maketitle

\begin{abstract}
We develop an approach to the deformation quantization on the real plane
with an arbitrary Poisson structure which based on Weyl symmetrically
ordered operator products. By using a polydifferential representation
for deformed coordinates $\hat x^j$ we are able to formulate a simple and
effective iterative procedure which allowed us to calculate the fourth
order star product (and may be extended to the fifth order at the expense
of tedious but otherwise straightforward calculations). Modulo some
cohomology issues which we do not consider here, the method gives an
explicit and physics-friendly description of the star products.
\end{abstract}

\section{Introduction}
The modern history of deformation quantization started with the paper
\cite{Bayen:1977ha}, see \cite{Dito:2002dr} for an overview.
An important ingredient of the deformation quantization program
is a construction of a star product. One takes the algebra
$\mathcal{A}$ of smooth functions on a Poisson manifold equipped
with a Poisson structure $\omega$ and deforms $\mathcal{A}$ to
$\mathcal{A}_\omega$ by introducing a star product in such a way that
the star commutator of any two functions mimics in the leading order
of the $\omega$-expansion the Poisson bracket between these functions.
The existence of deformation quantization of symplectic manifolds
was demonstrated in \cite{DWL}, and in \cite{OMY} it was shown that over
any symplectic manifold exists a Weyl manifold.
In 1997 Kontsevich \cite{Kontsevich:1997vb} demonstrated the existence of
a star product on any smooth Poisson manifold and presented a formula
which, in principle, allowed to calculate such a product. However,
if one needs the star product beyond the second order of the expansion
the formulae of Kontsevich are not very useful since there is no
regular method to calculate the integrals involved. Besides, there is
another (psychological) reason to look for a more simple formulation of
the star product. The machinery of the Formality Theorem is a bit too
heavy to be easily digested by a considerable part of the physics
community. Also note that the Formality Theorem actually gives much more
than just a star product. Therefore, when the hard part of the job is
already done, one can start looking for a formulation of star products
in a more physical language admitting a simpler computational algorithm  
for higher orders of the expansion of star products.

In this paper we develop an approach based on the symmetric ordering of
the operators $\hat x^j$ which represent deformed Cartesian coordinates
on $\mathbb{R}^N$. We require that $\hat x^j$ satisfy the main commutation
relation $[\hat x^j,\hat x^k]=2\alpha \hat \omega^{jk}(\hat x)$ (with $\alpha$
being a formal deformation parameter) and construct a $1$-polydifferential
representation for $\hat x^j$. The main advantage of this approach is
the existence of an iterative procedure in $\alpha$
such that the order $(n-1)$ in $\hat \omega$ allows to calculate the
order $n$ in $\hat x$, which in turn defines the order $n$ of
$\hat \omega$. Besides, we operate with objects which are familiar from
quantum mechanics. Some elements of this construction can be found
scattered in the literature. Behr and Sykora \cite{Behr:2003qc} used
the Weyl ordered operator products to construct generic star products
on $\mathbb{R}^N$, but without constructing a polydifferential
representation for $\hat x$. Technical difficulties did not allow them
to go beyond the second order of the expansion. Particular realization
of generators as formal power series is a very natural tool in the case
of linear Poisson structure (Lie algebras), 
which was effectively used to analyze star products 
\cite{Bordemann,Durov:2006iv,Meljanac:2006ui,Chrys,Meljanac:2008ud}.

Computations of the third order star product were done in \cite{Penkava}
by using the Hochschild cohomologies and in \cite{Zotov:2000ec} by changing
variables in the Moyal formula. Our results to this order are in perfect
agreement with \cite{Penkava}. Precise relations of our star product to
that of \cite{Zotov:2000ec} are less clear to us, but at least we do
not see any contradictions.

We like to stress, that in our procedure the calculations of higher
orders of the star product go without much intellectual effort,
except for solving the consistency condition at odd orders (for which
we cannot present any generic algorithm). However, in some cases
(e.g., two-dimensional real plane or linear Poisson structure)
the consistency condition is solved automatically, and calculations may
proceed up to an arbitrary high order of the expansion.

This paper is organized as follows. In the next section we consider
the Weyl ordered operator products and a star product. The iterative
procedure of calculations of the star product is formulated in section
\ref{sec-gen}, and actual calculations are contained in section
\ref{sec-calc}. Last section contains conclusions and discussion.
Some useful formulae are collected in the Appendix. 
\section{The Weyl map and the star product}\label{sec-map}
Consider a set of noncommuting operators $\hat x^j$, $j=1,\dots,N$ and
a function $f$ on $\mathbb{R}^N$ which can be expanded in the Taylor 
series around zero,
\begin{equation}
f(x)=\sum_{n=0}^\infty f^{(n)}_{i_1\dots i_n}x^{i_1}\dots x^{i_n}
\label{Tex}
\end{equation}
with $x^i$ being Cartesian coordinates on $\mathbb{R}^N$. We associate
to this function a symmetrically (Weyl) ordered operator function 
$\hat f(\hat x)$ according to the rule
\begin{equation}
\hat f(\hat x)=\sum_{n=0}^\infty f^{(n)}_{i_1\dots i_n}
\sum_{P_n} \frac 1{n!} P_n(\hat x^{i_1}\dots \hat x^{i_n}),\label{Weyl1}
\end{equation}
where $P_n$ are all permutations of $n$ elements. We shall also use another
notation,
\begin{equation}
\hat f \equiv W(f).\label{othernot}
\end{equation}

There is another, more convenient form of the Weyl ordering. Let $\tilde f(p)$
be a Fourier transform of $f$,
\begin{equation}
\tilde f(p) = \int d^Nx\, f(x)\, e^{ip_jx^j}.\label{Four}
\end{equation}
Then
\begin{equation}
\hat f\left( \hat{x}\right)  =\int \frac{d^{N}p}{\left( 2\pi
\right) ^{N}}\tilde{f}\left( p\right) e^{-ip_{m}\hat{x}^{m}}.  \label{2}
\end{equation}
(More on various orderings and corresponding integral representations
see in \cite{Agarwal:1971wc}).

Consider a skew-symmetric matrix-valued function $\omega^{ij}(x)$.
Let us now impose a commutation relation on the operators $\hat
x$:
\begin{equation}
[\hat x^i,\hat x^j]=2\alpha \hat \omega^{ij}(\hat x),\label{comx}
\end{equation}
where $\alpha$ is a formal expansion parameter. In the physics
literature $\alpha  =i\hbar /2$. This equation serves both as a constraint
on the star-product commutator and as a definition of the deformed
Poisson structure, order by order in $\alpha$, see below for details.
The commutation relation
(\ref{comx}) yields a consistency condition
\begin{equation}
[\hat x^i,\hat \omega^{jk}(\hat x)]+\mbox{cycl.}(ijk)=0.
\label{conscond}
\end{equation}

We shall look for a representation of the operators $\hat x$ 
as polydifferential operators in the form
of an $\alpha$-expansion,
\begin{equation}
\hat{x}^{i}=x^{i}+\overset{\infty }{\underset{n=1}{\sum }}\Gamma ^{i\left(
n\right) }\left( \alpha ,x\right) \left( \alpha \partial \right) ^{n}~,
\label{3}
\end{equation}%
where 
\begin{equation}
\Gamma ^{i\left( n\right) }\left( \alpha ,x\right) =\Gamma
^{ii_{1}...i_{n}}\left( \alpha ,x\right)
\end{equation}%
and each $\Gamma$, in turn, is expanded in $\alpha$
\begin{equation}
\Gamma ^{i\left( n\right) }\left( \alpha ,x\right) =\overset{\infty }{%
\underset{k=0}{\sum }}\alpha ^{k}\Gamma _{k}^{i\left( n\right) }\left(
x\right) ~.  \label{4}
\end{equation}%

We define a star product as
\begin{equation}
W(f\star g)=W(f)\cdot W(g).\label{starpr}
\end{equation}
This product is associative due to associativity of the operator
products. For a constant $\omega^{ij}$ the formula (\ref{starpr})
is very well known from quantum mechanics \cite{BerShub}. For a linear
Poisson structure this is in essence the Gutt's construction 
\cite{Gutt,Dito}. For arbitrary $\omega^{ij}(x)$ this star product
was used in \cite{Behr:2003qc}.

We also like to keep the unity of the algebra of
functions undeformed, yielding $f\star 1=f$, or
\begin{equation}
W(f)\cdot 1=f.\label{unity}
\end{equation}
The equations (\ref{starpr}) and (\ref{unity}) yield the following
formula
\begin{equation}
(f\star g)(x)=W\left( f\right) g(x)
=\hat f\left( \hat{x}\right) g(x)~,  \label{d3}
\end{equation}%
where the right hand side means an action of a polydifferential operator
on a function. 

Note that neither $x^j$ nor unit function are Schwarz class.
They do not belong to the algebra we are going to deform, but rather
to the algebra of multipliers (see \cite{GraciaBondia:2001ct}).
Of course, it is important that they stay in the multiplier
algebra also after the deformation. However, since we shall work
with formal expansions only, these subtleties will be ignored.

There is an obvious symmetry of this construction which changes the sign
on $\alpha$, $\alpha\to -\alpha$, and reverses the order of the operators.
(This is nothing else than the complex conjugation if one remembers that
$\alpha = i\hbar/2$ is, in fact, imaginary). This symmetry interchanges $f$
and $g$ in (\ref{starpr}). A consequence of this symmetry is that even
orders in $\alpha$ are symmetric with respect to interchanging $f$ and
$g$, while the odd orders are antisymmetric. Note, that only the deformations
of $\mathcal{A}$ which have this property are usually regarded as
deformation quantizations.

The matrix-valued function $\omega^{ij}(x)$ is also expanded in a power
series in $\alpha$, and the $\alpha^0$ term is a Poisson structure, i.e.,
$\omega_0^{ij}$ satisfies the Jacobi identity
\begin{equation}
\omega_0^{kl}\partial _{l}\omega_0^{ij}+\omega_0^{jl}\partial _{l}\omega_0
^{ki}+\omega_0^{il}\partial _{l}\omega_0^{kj}=0~.  \label{Jacobi}
\end{equation}
Since the coefficient functions $\Gamma^{i(n)}$ are contracted in
(\ref{3}) with partial derivatives only the part of $\Gamma^{i(n)}$
which is symmetric in the last $n$ indices contributes. From now on
we assume that $\Gamma^{i(n)}$ {\it is} symmetric in the last $n$ indices.

The star product will be constructed by using an iterative procedure.
One start with a zeroth order $\omega$ which may be an arbitrary Poisson
structure and solves (\ref{comx}) to the order $\alpha$ to obtain the
leading order of the coefficient functions $\Gamma$ and of the operators
$\hat x^j$. Then, by using these operators, one constructs a Weyl ordered
$\hat \omega^{ij}(\hat x)$ in terms of $\omega_0$, substitutes it to the
consistency condition (\ref{conscond}) and finds corrections to the
$\omega_0$. This corrections measure the failure of $\omega_0$ to
be a Poisson bivector. 
(Until rather high orders in $\alpha$ the condition (\ref{conscond})
is satisfied automatically and no corrections to $\omega$ appear).
Then one repeats the procedure with this new $\omega$ in the next order
in $\alpha$. Note, that higher order corrections to $\omega$ are expresed
through $\omega_0$, so that at the end we arrive at a star product
depending on the Poisson bivector $\omega_0$.

Note, that the existence of the representation of the star product we
are looking for is not obvious. Our procedure gives a constructive proof
of the existence which is valid to a certain order in $\alpha$ 
generically, and to all orders in the particular cases mentioned 
above\footnote{One can probably show generically the existence of the
Weyl representation of star product by combining the results of
\cite{Bordemann,BuWa}. We do not pursue this way since our emphasis
is on constructive aspects.}.

There is aan alterantive set of notations invented in \cite{Cargo}
for the Moyal product. Using this set may simplify the formulae to
some extent. We, however, stick to our notations to avoid misprints.
\section{General properties of the expansion}\label{sec-gen}
\subsection{Expansion of $\hat x$}
Consider the restrictions on the coefficient functions $\Gamma$
imposed by the condition (\ref{unity}). This condition is equivalent
to the requirement
\begin{equation}
W(x^{i_1} \dots x^{i_n})\cdot 1=x^{i_1} \dots x^{i_n}
\label{unit2}
\end{equation}
for all monomials of the coordinates and in all orders in $\alpha$.
Note, that the right hand side of (\ref{unit2}) does not depend on
$\alpha$. Consequently, all contributions to the left hand side
higher than zeroth order in $\alpha$ must vanish. It is easy to see,
that this, in turn, is equivalent to vanishing of totally symmetrized
parts of all $\Gamma$,
\begin{equation}
\Gamma ^{(ii_{1}...i_{k})}=0~.  \label{d4}
\end{equation}
Or, the contraction of each $\Gamma^{(n)}$ with $n+1$ commuting
vectors $p_i$ vanishes,
\begin{equation}
p_{i}p_{i_{1}}...p_{i_{k}}\Gamma ^{ii_{1}...i_{k}}=0~,\label{tr2}
\end{equation}
For brevity, by somewhat stretching the terminology, we shall call this
condition the tracelessness condition.

Let us now turn to the consequences of the commutation relation 
(\ref{comx}). As we shall see below, the relation (\ref{comx})
allows to express
\begin{equation}
\Gamma_k^{[ii_1]i_2\dots i_p}\equiv
\Gamma_k^{ii_1i_2\dots i_p}-
\Gamma_k^{i_1ii_2\dots i_p}\label{Gasym}
\end{equation}
with $k+p=n$ in terms of the lower order $\Gamma$'s.
It is convenient to expand the right hand side of
(\ref{comx}) as
\begin{equation}
\hat \omega^{ij}(\hat x)= \hat \omega_n^{ij}
+o(\alpha^n)\,.\label{expom}
\end{equation}
where $\hat \omega_n$ is of the order $\alpha^n$. We also introduce
corresponding expansions for $\hat x$ as
\begin{equation}
\hat x^j=\hat x^j_n +o(\alpha^{n})\,,\qquad
\hat x^j_{n+1}=\hat x^j_n +\alpha^{n+1}\gamma_{n+1}^j\,.
\label{expx}
\end{equation}
Suppose, one has already calculated the expansion of
$\hat x$ up to the order $\alpha^{n}$, i.e
\begin{equation}
[\hat x_n^i,\hat x_n^j]=2\alpha \hat \omega_{n-1} +o(\alpha^n).
\label{coman}
\end{equation}
Next we check the consistency condition (\ref{conscond}) in the order
$\alpha^n$:
\begin{equation}
\left[ \hat{x}_{n}^{k},\hat{\omega}_{n}^{ij}\right] +\left[ \hat{x}_{n}^{j},%
\hat{\omega}_{n}^{ki}\right] +\left[ \hat{x}_{n}^{i},\hat{\omega}_{n}^{jk}%
\right] =o\left( \alpha ^{n}\right) ~.  \label{b4}
\end{equation}
To the lowest orders this equation is satisfied automatically for any
Poisson bivector $\omega^{ij}$. In the higher orders it does not, and
(\ref{b4}) must be considered as a condition on non-Poisson corrections
to $\omega^{ij}$. Solving this equation
is a rather nontrivial part of the procedure. This part will be considered
separately in sec.\ \ref{sec-om}. For the time being we assume that 
corresponding corrections are constructed and (\ref{b4}) is satisfied.

In order to construct the next, $(n+1)$th order in the 
decomposition we have to solve the
following equation
\begin{equation}
\left[ \hat{x}_{n+1}^{i},\hat{x}_{n+1}^{j}\right] =2\alpha \hat{\omega}%
_{n}^{ij}+o\left( \alpha ^{n+1}\right) ~.  \label{b5}
\end{equation}%
Or, 
\begin{equation*}
\left[ \hat{x}_{n}^{i}+\alpha ^{n+1}\gamma _{n+1}^{i},\hat{x}_{n}^{j}+\alpha
^{n+1}\gamma _{n+1}^{j}\right]  =2\alpha \hat{\omega}_{n}^{ij}+o\left(
\alpha ^{n+1}\right) \end{equation*}
which implies
\begin{equation}
\left[ \hat{x}_{n}^{i},\hat{x}_{n}^{j}\right] +\alpha ^{n+1}\gamma
_{n+1}^{[ij]} =2\alpha \hat{\omega}_{n}^{ij}+o\left( \alpha ^{n+1}\right) ,
\label{408}
\end{equation}
where
\begin{equation}
\gamma^{[ij]}_{n+1}\equiv \sum_{p+k=n+1} p
\Gamma_k^{[ij]i_2\dots i_p}\partial_{i_2}\dots \partial_{i_p}\,.
\label{gamij}
\end{equation}

Therefore, the antisymmetric part of $\Gamma_{n+1}$ is determined from
the equation
\begin{equation}
\alpha ^{n+1}\gamma _{n+1}^{[ij]}=G_{n+1}^{ij}+o\left( \alpha ^{n+1}\right) ,
\label{GGam}
\end{equation}
where%
\begin{equation}
G_{n+1}^{ij}=2\alpha \hat{\omega}_{n}^{ij}-\left[ \hat{x}_{n}^{i},\hat{x}%
_{n}^{j}\right] +o\left( \alpha ^{n+1}\right) ~.  \label{bb}
\end{equation}%
From this equation $G_{n+1}^{ij}$ is defined up to terms $o(\alpha^{n+1})$.
We do not include any higher order terms in $G_{n+1}^{ij}$, so that
 $G_{n+1}^{ij}\sim \alpha^{n+1}$.
The operators $G_{n+1}^{ij}$ can be expanded as
\begin{equation}
G_{n+1}^{ij} = \sum_p G_{n+1}^{iji_2\dots i_p}\partial_{i_2}\dots
\partial_{i_p}\,.\label{expG}
\end{equation}
The coefficient functions in this expansion are antisymmetric in $i,j$
and symmetric in $i_2,\dots,i_p$
by the construction, and they also have the following property.
\begin{lemma}\label{L1} The functions $ G_{n+1}^{iji_2\dots i_p}$ obey the
cyclic condition
\begin{equation}
 G_{n+1}^{iji_2\dots i_p}+{\rm cycl.}(iji_2)=0.\label{cyclG}
\end{equation}
\end{lemma}
\begin{proof}
From (\ref{b4}) one has
\begin{equation}
\left[ \hat{x}_{n}^{k},2\alpha \hat{\omega}_{n}^{ij}\right] 
+\mbox{cycl.}(kij)=o\left(
\alpha ^{n+1}\right) ~,
\end{equation}%
or, using (\ref{bb}),
\begin{equation}
\left[ \hat{x}_{n}^{k},G_{n+1}^{ij}+\left[ \hat{x}_{n}^{i},\hat{x}_{n}^{j}%
\right] \right] +\mbox{cycl.}(kij)=o\left( \alpha ^{n+1}\right) ~.  \label{b6}
\end{equation}
The equation 
\begin{equation}
\left[ \hat{x}_{n}^{k},\left[ \hat{x}_{n}^{i},\hat{x}_{n}^{j}\right] \right]
+\mbox{cycl.}(kij)=0
\end{equation}
holds true at all orders of $\alpha $, including the order $\alpha ^{n+1}$.
Therefore, from (\ref{b6}) one obtains
\begin{equation}
\left[ x_{n}^{k},G_{n+1}^{ij}\right] +cycl=0~.  \label{b9}
\end{equation}
Next one substitutes the expansion (\ref{expG}) in (\ref{b9}),
calculates the commutator, and uses the symmetry of $G_{n+1}$
in the last $p-1$ indices to complete the proof.
\end{proof}

Because of the symmetry of $G_{n+1}^{iji_2\dots i_p}$ in the last $p-1$
indices, the cyclic conditions holds for permutations of $(i,j,i_k)$
for any $k=2,\dots,p$.

Now we are able to construct $\Gamma_p^{ij\dots k}$ and thus the
operator $\hat x$ to the order $\alpha^{n+1}$ by symmetrising
the functions $G_{n+1}^{iji_1\dots i_k}$. The equations which should
be satisfied by the functions $\Gamma$ in order to solve the commutation
relation (\ref{b5}) read
\begin{equation}
G_{n+1}^{ji_1\dots i_p}=\alpha^{n+1} p \Gamma_k^{[ji_1]\dots i_p},
\qquad k+p=n+1.\label{GaG}
\end{equation}
A solution to these equations is given by the following Lemma.
\begin{lemma}\label{L2}
The tensors 
\begin{equation}
\Gamma^{ji_1\dots i_p}_k=\frac {\alpha^{-(n+1)}}{p(p+1)}
\left( G_{n+1}^{ji_1i_2\dots i_p}+
G_{n+1}^{ji_2i_1i_3\dots i_p}+\dots 
+G_{n+1}^{ji_pi_1i_2\dots i_{p-1}} \right) \label{GamG}
\end{equation}
are symmetric in the last $p$ indices,
satisfy the equation (\ref{GaG}) and the tracelessness
condition (\ref{d4}).
\end{lemma}
\begin{proof} The symmetry follows by the construction, and the tracelessness
is a consequence of the antisymmetry of $G_{n+1}^{ji_1i_2\dots i_p}$ in the
first two indices. To prove the remaining assertion, let us consider 
antisymmetrized in $j$ and $i_1$ combinations of the tensors\footnote{The
word tensor here is just a short hand notation for an object with many
indices. No particular transformation properties with respect to any
group are assumed} 
entering
the right hand side of (\ref{GaG}).
\begin{equation*}
G_{n+1}^{ji_1i_2\dots i_p}-
G_{n+1}^{i_1ji_2\dots i_p}=
2 G_{n+1}^{ji_1i_2\dots i_p}
\end{equation*}
due to the antisymmetry of $G$ in the first two indices.
\begin{equation*}
G_{n+1}^{ji_2i_1\dots i_p}-G_{n+1}^{i_1i_2j\dots i_p}=
 G_{n+1}^{ji_1i_2\dots i_p}\end{equation*}
due to the cyclic condition in $j,i_1,i_2$. Remaining $p-2$ combinations
are treated similarly, and the assertion follows immediately.
\end{proof}

This Lemma implies that the tensors (\ref{GamG}) are indeed the coefficient
functions of the expansion of $\hat x^j$ we are looking for.

For the notational convenience we define $\star_k$ which is
a part of the star product having the order $\alpha^k$. 
\begin{equation}
f\star g =\sum_{k=0}^n f\star_k g +O(\alpha^{n+1}).
\label{stark}
\end{equation}

To evaluate the star product and $\hat \omega (\hat x)$ to a given
order of $\alpha$ we shall need an effective tool to calculate an
$\alpha$ expansion of the Weyl-ordered operator $\hat f(\hat x)$ for
any given $f$ and a given expansion of $\hat x$. To this end we shall
use the integral representation (\ref{2}) and the Duhamel formula
\begin{equation}
e^{A+B}=e^{A}+\overset{1}{\underset{0}{\int }}e^{\left( A+B\right)
s}Be^{\left( 1-s\right) A}ds~.  \label{8}
\end{equation}%
Here $A+B=-ip_{i}\hat{x}^{i},\ A=-ip_{i}x^{i}$ and $B=-ip_{i}\left( \hat{x}
^{i}-x^{i}\right)$. Therefore, $B$ is of the order $\alpha^1$, and
$A$ is of the order $\alpha^0$, but, because of the property (\ref{tr2}),
each commutator is at least one order higher, 
$\left[ B,A\right] =O(\alpha ^{2}),\ \left[ \left[ B,A%
\right] ,A\right] =O(\alpha ^{3}),\ \left[ \left[ \left[ B,A\right] ,A\right]
,A\right] =O(\alpha ^{4})$. By using these rules, one easily finds,
\begin{eqnarray}
e^{A+B} &=&e^{A}\left(1 +B+\frac{1}{2}\left[ B,A\right] +\frac{1}{2}%
B^{2} \right. \label{dec} \\
&&+\frac{1}{6}\left[ \left[ B,A\right] ,A\right] +\frac{1}{3}\left[
B,A\right] B+\frac{1}{6}B\left[ B,A\right] +\frac{1}{6}B^{3}
\notag \\
&&+\frac{1}{24}\left[ \left[ \left[ B,A\right] ,A\right] ,A\right] +%
\frac{1}{8}\left[ \left[ B,A\right] ,A\right] B+\frac{1}{8}\left[
B,A\right] ^{2}  \notag \\
&&+\frac{1}{24}B\left[ \left[ B,A\right] ,A\right] +\frac{1}{8}
\left[ B,A\right] B^{2}+\frac{1}{12}B\left[ B,A\right] B  \notag \\
&&\left.
+\frac{1}{24}B^{2}\left[ B,A\right] +\frac{1}{24}B^{4}\right)
+O\left(
\alpha ^{5}\right) ~.  \notag
\end{eqnarray}

The general strategy of calculations of the $\alpha$-expansion of the
star product is as follows. Starting with an $n$th order operators
$\hat x^j$ one calculates the operator $\hat \omega^{ij}(\hat x_n)$ and
the operators $G^{ij}_{n+1}$ from the formula (\ref{bb}). The next order
operators are the constructed by using Lemma \ref{L2}. Then one has to check
the consistency condition at this order and calculate corrections to
$\omega$, if needed (see sec.\ \ref{sec-om}). Apart from the corrections 
to $\omega$, the procedure goes automatically and is even suitable for
a computer. 
\subsection{Corrections to $\omega$}\label{sec-om}
Here we give a short overview of what happens if corrections
to $\omega$ are needed. A more detailed discussion can be found
in sec.\ \ref{sec-cor3} below where we deal with a particular case
of the third order star product. Suppose we have completed our
construction at some order $n$, i.e., we have operators $\hat x^j_n$
fulfilling the relation (\ref{coman}). Next, we take the function
$\omega_{n-1}$ and construct the corresponding Weyl ordered
operator $\hat \omega^{ij}(\hat x_n)$ and the star product 
$\tilde \star_n$, where the twiddle means that we still have to
check the consistency condition (\ref{b4}) at the order $n$.
Let us assume for simplicity that at all lower orders no
corrections to $\omega$ appeared, i.e., that all lower order
consistency conditions were satisfied by 
$\omega^{ij}=\omega_0^{ij}$. Let us denote 
$\tilde \star\equiv \star_1 +\dots +\star_{n-1}+\tilde \star_n$
and compute
\begin{equation}
x^i\tilde \star \omega^{jk}-\omega^{jk}\tilde \star x^i +
\mbox {cycl}(ijk)\,.
\label{tilcyc}
\end{equation}
If the expression (\ref{tilcyc}) is $o(\alpha^n)$ for $\omega=\omega_0$,
no corrections are needed at this order as well. If the expression
(\ref{tilcyc}) contains some $\alpha^n$ terms (lower order terms vanish
due to the lower order consistency conditions), we have to correct
$\omega$ and, consequently, the star product. It is easy to see, that
no correction to $\omega$ at the order $\alpha^n$ will do the job.
We must correct $\omega$ in the order $\alpha^{n-1}$, so that
$\omega=\omega_0+\alpha^{n-1}\omega_{n-1}$. This looks very dangerous
for the whole approach, since we are going to correct the order we
have already constructed. However, this is not so. Let us consider
more closely what is going on. The only effect the correction to
$\omega$ at the order $\alpha^{n-1}$ has on the operators $\hat x_n$
is that now the part $\Gamma_{n-1}^{ij}$ is non-zero,
\begin{equation}
\Gamma_{n-1}^{ij}=\omega_{n-1}^{ij}.\label{Gn1}
\end{equation}
The star product is modified in the $n$-th order only,
$\tilde \star_n\to \star_n$,
\begin{equation}
f\star_n g =f\tilde \star_n g +\alpha^n \omega^{ij}_{n-1}
\partial_i f \partial_j g.\label{fgfg}
\end{equation}

This formula is related to the well known ambiguity in the star products
(see, e.g., \cite{Penkava}). If the star product $\tilde \star$ is associative
up to $\alpha^{n+1}$ terms, then the star product $\star$ is also associative
up to $\alpha^{n+1}$ independently of the choice of $\omega_{n-1}^{ij}$.
Therefore, both products are legitimate star products. The only difference
is that $\tilde \star$ cannot be extended to the next order 
{\it with our methods}, while $\star$ can. If we need a star product to the
order $n$ only, we can as well ignore all corrections coming from the
$n$th and higher order consistency conditions.

Another problem is to find actually an expression for $\omega_{n-1}^{ij}$
which solves the consistency condition. Here we shall not attempt to
present any method of solving the consistency condition for
$\omega_{n-1}^{ij}$ or even analyze the existence of a solution for the
following reasons. The equation for $\omega_{n-1}^{ij}$ appearing in our 
approach are practically identical to that coming from the Hochschild
cohomologies \cite{Penkava}. Studying relations between our operator
algebra approach and Hochschild cohomologies is an interesting problem
on its own right, and we are going to address it in a separate work.
For practical purposes, to calculate the star product up to the fifth
order, one only needs to solve a single non-trivial consistency
condition at the third order. This can be done directly, see sec.\
\ref{sec-cor3}. Besides, third order consistency conditions have been
already analyzed in \cite{Penkava} in a different formalism.  
\section{Calculation of the star product}\label{sec-calc}
\subsection{First order star product}
As a warm up, let us calculate the zeroth and  first
 orders of the $\alpha$-expansion
of the star product. In the zeroth order, the condition (\ref{comx})
reads $[\hat x^i,\hat x^j]=0$, so that we have an undeformed commutative
algebra, $\hat x^j=x^j$, and, as expected,
\begin{equation}
f\star_0 g=f\cdot g. \label{zerost}
\end{equation}

To the first order in $\alpha$ the expansion (\ref{3})
reads
\begin{equation}
\hat{x}^{i}=x^{i}+\alpha \Gamma _{0}^{ij}\left( x\right) \partial
_{j}+O\left( \alpha ^{2}\right) ~.  \label{x1}
\end{equation}%
By substituting this expansion in (\ref{comx}) and keeping only the
terms which are linear in $\alpha$, we obtain
\begin{equation}
\Gamma _{0}^{\left[ ij\right] }
=\Gamma _{0}^{ij}\left( x\right) -\Gamma _{0}^{ji}\left( x\right) =2\omega
^{ij}\left( x\right) ~,  \label{6}
\end{equation}%
which implies
\begin{equation}
\Gamma _{0}^{ij}\left( x\right) =\omega ^{ij}\left( x\right)
+S_{0}^{ij}\left( x\right) ~.  \label{7}
\end{equation}%
Note, that to this order in $\alpha$ the Weyl ordering is trivial,
$\hat \omega (\hat x)=\omega (x)$.
The symmetric part $S^{ij}_0$ is eliminated by the tracelessness condition
(\ref{d4}), $S^{ij}_0=0$, and
\begin{equation}
\Gamma _{0}^{ij}\left( x\right) =\omega ^{ij}\left( x\right) ~.
\label{Gamzer}
\end{equation}

The Duhamel formula (\ref{8}) gives
\begin{eqnarray}
&&e^{-ip_{m}\hat{x}^{m}}=e^{A}+e^{A}B+O\left( \alpha ^{2}\right) 
\label{10} \\
&&\qquad\quad =e^{-ip_{m}x^{m}}
-i\alpha e^{-ip_{m}x^{m}}p_{j}\omega ^{jk}\left( x\right)
\partial _{k}+O\left( \alpha ^{2}\right) ~.  \notag
\end{eqnarray}%
By using the equation (\ref{d3}) we immediately obtain
\begin{equation}
f\star_1 g=\alpha \omega^{ij}\partial_i f\,\partial_j g,\label{onest}
\end{equation}
and the expansion
\begin{equation}
\hat \omega ^{ij}\left( \hat{x}\right) =\omega ^{ij}\left( x\right) +\alpha
\partial _{l}\omega ^{ij}\left( x\right) \omega ^{lk}\left( x\right)
\partial _{k}+O\left( \alpha ^{2}\right) ~,  \label{11}
\end{equation}%
which will be used below to calculate the next order star product.

The consistency condition (\ref{conscond}) to this order 
\begin{equation}
(x^k\star_1 \omega^{ij} - \omega^{ij}\star_1 x^k)+
\mbox{cycl.}(kil)=0\label{cc1}
\end{equation}
yields
the Jacobi identity (\ref{Jacobi}), i.e., $\omega = \omega_0$ is
a Poisson bivector. We shall drop the subscript $0$ from the notations
whenever this cannot lead to a confusion.
\subsection{Second order star product}\label{sec-2nd}
As is seen from the equation (\ref{11}), at the order $\alpha^1$ the
operator $\hat \omega^{ij}$ is a first order differential operator with
a vanishing zeroth order part. Consequently, $\gamma_2^i$ does not have
a first order part, and we may write
\begin{equation}
\hat{x}_2^{i} =x^{i}+\alpha \omega^{ij}\left( x\right) \partial
_{j}+\alpha ^{2}\Gamma _{0}^{ijk}\left( x\right) \partial _{j}\partial _{k}
\label{5}.\end{equation}
The commutation relation (\ref{coman}) with $n=1$ yields
\begin{equation}
2\alpha^2 \Gamma _{0}^{[il]k}=G_2^{ilk}=
2\alpha^2 \left( \omega ^{jk}\partial _{j}\omega ^{il}+%
\frac{1}{2}\omega ^{lj}\partial _{j}\omega ^{ik}-\frac{1}{2}\omega
^{ij}\partial _{j}\omega ^{lk}\right).\label{13}
\end{equation}
The cyclic condition 
\begin{equation}
G_{2}^{ilk}+G_{2}^{kil}+G _{2}^{lki}=0 \label{cycl2}
\end{equation}
can easily be checked. It is equivalent to the Jacobi identity on
$\omega^{ij}$. With the help of this identity we can rewrite (\ref{13}) as%
\begin{equation}
G_{2}^{ilk}=\alpha^2 \omega
^{jk}\partial _{j}\omega ^{il}~.  \label{15}
\end{equation}

According to the general prescription of Lemma \ref{L2},
\begin{equation*}
\alpha^2 \Gamma_0^{ilk}=\frac 16 \left( G_2^{ilk}+G_2^{ikl}\right),
\end{equation*}
or,
\begin{equation}
\Gamma _{0}^{ilk}=\frac{1}{6}\omega ^{jk}\partial _{j}\omega ^{il}+\frac{1}{6%
}\omega ^{jl}\partial _{j}\omega ^{ik}~.  \label{15c}
\end{equation}%

Obviously, $\Gamma _{0}^{ilk}$ obeys
the tracelessness condition (\ref{d4}). Eq.\ (\ref{13}) can be checked 
directly by using the Jacobi identity.

Now we are ready calculate star product up to the second 
order using the formula (\ref{d3}). First we calculate
\begin{equation*}
[B,A]=-\frac{\alpha ^{2}}{3}p_{i}p_{k}\omega ^{ji}\partial _{j}\omega
^{kl}\partial _{l}+O\left( \alpha ^{3}\right).\end{equation*}
The decomposition (\ref{dec}) yields
\begin{eqnarray}
&&e^{-ip_{i}\hat{x}^{i}}=e^{-ip_{m}x^{m}}-i\alpha
e^{-ip_{m}x^{m}}p_{i}\omega ^{ij}\left( x\right) \partial _{j}  \notag \\
&&-\frac{\alpha ^{2}}{2}e^{-ip_{m}x^{m}}p_{i}p_{k}\omega ^{ij}\omega
^{kl}\partial _{j}\partial _{l}-\frac{\alpha ^{2}}{3}%
e^{-ip_{m}x^{m}}p_{i}p_{k}\omega ^{ij}\partial _{j}\omega ^{kl}\partial _{l}
\label{17} \\
&&+\frac{i\alpha ^{2}}{6}e^{-ip_{m}x^{m}}p_{k}\left( \omega ^{jk}\partial
_{j}\omega ^{il}+\omega ^{jl}\partial _{j}\omega ^{ik}\right) \partial
_{i}\partial _{l}+O\left( \alpha ^{3}\right) ~.  \notag
\end{eqnarray}%
From eq.\ (\ref{2}), we have
\begin{eqnarray}
&&f\left( \hat{x}\right) =f(x)+\alpha \omega ^{ij}\partial _{i}f\partial
_{j}  \notag \\
&&\qquad +\frac{\alpha ^{2}}{2}\omega ^{ij}\omega ^{kl}\partial _{i}\partial
_{k}f\partial _{j}\partial _{l}+\frac{\alpha ^{2}}{3}\omega ^{ij}\partial
_{j}\omega ^{kl}\partial _{i}\partial _{k}f\partial _{l}  \label{18} \\
&&\qquad -\frac{\alpha ^{2}}{3}\left( 
\omega ^{jk}\partial _{j}\omega ^{il}+\omega
^{jl}\partial _{j}\omega ^{ik}\right) \partial _{k}f\partial _{i}\partial
_{l}+O\left( \alpha ^{3}\right) .  \notag
\end{eqnarray}%
And then, from (\ref{d3}) we obtain
\begin{eqnarray}
&&(f\star g)(x)=\hat f\left( \hat{x}\right) g(x)=fg+\alpha \partial _{i}f\omega
^{ij}\partial _{j}g  \notag \\
&&+\frac{\alpha ^{2}}{2}\omega ^{ij}\omega ^{kl}\partial _{i}\partial
_{k}f\partial _{j}\partial _{l}g+\frac{\alpha ^{2}}{3}\omega ^{ij}\partial
_{j}\omega ^{kl}\left( \partial _{i}\partial _{k}f\partial _{l}g-\partial
_{k}f\partial _{i}\partial _{l}g\right) +O\left( \alpha ^{3}\right) ~.
\label{19}
\end{eqnarray}
This expression coincides with the Kontsevich formula 
\cite{Kontsevich:1997vb}. The same expression was re-derived 
from the Weyl-ordered operator products by Behr and Sykora
\cite{Behr:2003qc}. Since the product (\ref{19}) coincides 
with known ones, there is no need to check the consistency
conditions (\ref{conscond}) at this order. The consistency condition
also follows from a more strong statement
\begin{equation}
f\star_2 g - g\star_2 f=0,\label{fggf2}
\end{equation}
which is a consequence of the symmetry we described in sec.\ \ref{sec-map}
and may be easily verified from (\ref{19}).
No correction to
$\omega$ appears, meaning that to this order 
$\omega^{ij}=\omega_0^{ij}$ is a Poisson structure. 

As a preparation to the next order calculations we also write
an expansion for $\hat\omega$:
\begin{eqnarray}
&&\hat\omega ^{ij}\left( \hat{x}\right) =\omega ^{ij}+\alpha \omega
^{kl}\partial _{k}\omega ^{ij}\partial _{l}  \notag \\
&&\qquad +\frac{\alpha ^{2}}{2}\omega ^{nk}\omega ^{ml}\partial _{n}\partial
_{m}\omega ^{ij}\partial _{k}\partial _{l}+\frac{\alpha ^{2}}{3}\omega
^{nk}\partial _{k}\omega ^{ml}\partial _{n}\partial _{m}\omega ^{ij}\partial
_{l}  \label{20} \\
&&\qquad 
-\frac{\alpha ^{2}}{3}\left( \omega ^{nk}\partial _{n}\omega ^{lm}+\omega
^{nl}\partial _{n}\omega ^{km}\right) \partial _{m}\omega ^{ij}\partial
_{k}\partial _{l}+O\left( \alpha ^{3}\right) ~.  \notag
\end{eqnarray}%
\subsection{Third order star product}
The operator $\hat \omega$ in the order $\alpha^2$, eq.\ (\ref{20}),
as well as in the previous order, does not contain a part which is a zeroth
order differential operator. Consequently, $\gamma_3^j$ does not contain
zeroth or first order terms, and we may write
\begin{eqnarray}
&&\hat x^i_2+\gamma_3^i=
\hat{x}^{i}_3=x^{i}+\alpha \Gamma _{0}^{ij}\left( x\right) \partial
_{j}+\alpha ^{2}\Gamma _{0}^{ijk}\left( x\right) \partial _{j}\partial _{k}+
\notag \\
&&\alpha ^{3}\left( \Gamma _{1}^{ijk}\left( x\right) \partial _{j}\partial
_{k}+\Gamma _{0}^{ijkl}\left( x\right) \partial _{j}\partial _{k}\partial
_{l}\right) +O\left( \alpha ^{4}\right) ~.  \label{21}
\end{eqnarray}%
The function $\Gamma^{ijk}_0$ is known, while $\Gamma_1^{ijk}$ and
$\Gamma_0^{ijkl}$ have to be defined. By comparing (\ref{20}) to the
commutator of (\ref{21}) we obtain (cf.\ (\ref{bb}))
\begin{equation}
G_{3}^{ijk}=\frac{\alpha^3}{3}\omega ^{nl}\left( 2\partial _{l}\omega
^{mk}\partial _{n}\partial _{m}\omega ^{ij}+\partial _{l}\omega
^{mj}\partial _{n}\partial _{m}\omega ^{ik}+\partial _{l}\omega
^{mi}\partial _{n}\partial _{m}\omega ^{kj}\right)   \label{23}
\end{equation}
and
\begin{eqnarray}
&&G_{3}^{ijkl}=\alpha^3 \left( \omega ^{nk}\omega ^{ml}\partial
_{n}\partial _{m}\omega ^{ij}+\frac{1}{3}\omega ^{kn}\partial _{n}\omega
^{lm}\partial _{m}\omega ^{ij}\right.  \notag \\
&&\qquad + \frac{1}{3}\omega ^{ln}\partial _{n}\omega ^{km}\partial _{m}\omega
^{ij}+\omega ^{jm}\partial _{m}\Gamma _{0}^{ikl}-\omega ^{im}\partial
_{m}\Gamma _{0}^{jkl}  \label{25} \\
&&\qquad +\left. \Gamma _{0}^{jkm}\partial _{m}\omega ^{il}+\Gamma
_{0}^{jlm}\partial _{m}\omega ^{ik}-\Gamma _{0}^{ikm}\partial _{m}\omega
^{jl}-\Gamma _{0}^{ilm}\partial _{m}\omega ^{jk}\right) ~.  \notag
\end{eqnarray}

The cyclic identities (Lemma \ref{L1}) can be verified directly as
consequences of the Jacobi identity. According to Lemma \ref{L2},
we have
\begin{equation}
\Gamma _{1}^{ijk}=\frac{1}{6}\omega ^{nl}\partial _{l}\omega ^{mk}\partial
_{n}\partial _{m}\omega ^{ij}+\frac{1}{6}\omega ^{nl}\partial _{l}\omega
^{mj}\partial _{n}\partial _{m}\omega ^{ik}.  \label{27}
\end{equation}
and
\begin{equation}
\alpha^3 \Gamma _{0}^{ijkl}=\frac{1}{12}\left( G _{3}^{ijkl}+
G_3^{ikjl}+G_{3}^{ilkj}\right) ~.  \label{32}
\end{equation}

With the help of the commutators (\ref{AB2}), the Duhamel formula 
(\ref{dec}), and the expressions (\ref{d3}), (\ref{23}) - (\ref{32})
we compute the third order star product.
\begin{eqnarray}
f\tilde \star_3 g&=
&\alpha^3 \left[ 
\frac{1}{3}\omega ^{nl}\partial _{l}\omega ^{mk}\partial _{n}\partial
_{m}\omega ^{ij}\left( \partial _{i}f\partial _{j}\partial _{k}g-\partial
_{i}g\partial _{j}\partial _{k}f\right) \right.  \notag \\
&&+\frac{1}{6}\omega ^{nk}\partial _{n}\omega ^{jm}\partial _{m}\omega
^{il}\left( \partial _{i}\partial _{j}f\partial _{k}\partial _{l}g-\partial
_{i}\partial _{j}g\partial _{k}\partial _{l}f\right)   \notag \\
&&+\frac{1}{3}\omega ^{ln}\partial _{l}\omega ^{jm}\omega ^{ik}\left(
\partial _{i}\partial _{j}f\partial _{k}\partial _{n}\partial _{m}g-\partial
_{i}\partial _{j}g\partial _{k}\partial _{n}\partial _{m}f\right) 
\label{37} \\
&&+\frac{1}{6}\omega ^{jl}\omega ^{im}\omega ^{kn}\partial _{i}\partial
_{j}\partial _{k}f\partial _{l}\partial _{n}\partial _{m}g  \notag \\
&&\left. +\frac{1}{6}\omega ^{nk}\omega ^{ml}\partial _{n}\partial _{m}\omega
^{ij}\left( \partial _{i}f\partial _{j}\partial _{k}\partial _{l}g-\partial
_{i}g\partial _{j}\partial _{k}\partial _{l}f\right)\right] ~.  \notag
\end{eqnarray}
We put a twiddle over the star to stress that (\ref{37}) does not 
include corrections to $\omega$ required by the consistency condition
(\ref{b4}) at the third order. However, this is a fully legitimate
star product. It only cannot be extended to the fourth order in our
procedure. Eq.\ (\ref{37}) is in agreement with \cite{Penkava} and does
fulfill the requirements following from associativity \cite{Zotov:2000ec}.
It does not coincide with a particular star product constructed in
 \cite{Zotov:2000ec} by changing coordinates in the Moyal formula, but
the difference presumably resides in the ambiguity discussed in 
sec.\ \ref{sec-om} plus a gauge transformation.
\subsection{Corrections to the third order star product}\label{sec-cor3}
Let us study the third order consistency conditions for the star product
we have obtained in the previous subsection. We have to calculate 
(\ref{tilcyc}) for $n=3$ and check whether $\alpha^3$ terms vanish.
This boils down to the condition
\begin{eqnarray}
&&x^{a}\tilde \star _{3}\omega ^{bc}-\omega ^{bc}
\tilde \star _{3}x^{a}+\mbox{cycl}(abc)
\label{38} \\
&&=\alpha^3 \left(
\frac{2}{3}\omega ^{nl}\partial _{l}\omega ^{mk}\partial _{n}\partial
_{m}\omega ^{aj}\partial _{j}\partial _{k}\omega ^{bc}+\frac{1}{3}\omega
^{nk}\omega ^{ml}\partial _{n}\partial _{m}\omega ^{aj}\partial _{j}\partial
_{k}\partial _{l}\omega ^{bc}\right)\nonumber\\
&&\qquad\qquad\qquad\qquad +\mbox{cycl}(abc)=0~.  \notag
\end{eqnarray}
The condition (\ref{38}) is not satisfied generically for $\omega^{bc}=
\omega_0^{bc}$, i.e. it does not follow from the Jacobi identity
\cite{Penkava}. Therefore, a correction to $\omega_0$ is needed, and,
according to sec.\ \ref{sec-om}, this has to be an $\alpha^2$ correction,
\begin{equation}
\omega ^{bc}\left( x\right) =\omega _{0}^{bc}\left( x\right) 
+\alpha ^{2}\omega _{2}^{bc}\left( x\right) +...
\label{39}
\end{equation}
This leads to a non-zero $\Gamma _{2}^{ij}$ according to the equation
\begin{equation}
\Gamma _{2}^{ij}-\Gamma _{2}^{ji}=2\omega _{2}^{ij}~,
\end{equation}
which, together with the tracelessness condition, yields
\begin{equation}
\Gamma _{2}^{ij}=\omega _{2}^{ij}~.  \label{40}
\end{equation}
The third order star product is changed, $\tilde\star_3\to \star_3$,
as
\begin{equation}
f{\star}_{3}g=f\tilde \star _{3}g+\alpha ^{3}\partial _{i}f\omega
_{2}^{ij}\partial _{j}g~.  \label{41}
\end{equation}
$\omega_2^{ij}$ has to be defined from the consistency condition which
reads
\begin{eqnarray}
&&x^{a}\star_{3}\omega_0^{bc}-\omega_0^{bc}
\star_{3}x^{a}+\alpha^2\,x^{a}\star _{1}\omega _{2}^{bc}
-\alpha^2\, \omega _{2}^{bc}\star
_{1}x^{a}+\mbox{cycl.}(abc)   \notag \\
&&=\alpha^3\left[ 
2\omega _{0}^{ad}\partial _{d}\omega _{2}^{bc}+2\omega _{2}^{ad}\partial
_{d}\omega _{0}^{bc}+\frac{2}{3}\omega _{0}^{nl}\partial _{l}\omega
_{0}^{mk}\partial _{n}\partial _{m}\omega _{0}^{aj}\partial _{j}\partial
_{k}\omega _{0}^{bc}\right.   \label{42} \\
&&\left.
+\frac{1}{3}\omega _{0}^{nk}\omega _{0}^{ml}\partial _{n}\partial _{m}\omega
_{0}^{aj}\partial _{j}\partial _{k}\partial _{l}\omega
_{0}^{bc}\right]+\mbox{cycl.}(abc)=0~.   \notag
\end{eqnarray}
$\omega_2^{jk}$ which solves this equation must be a sum of monomials
each containing three $\omega_0$ and four derivatives. The most general
ansatz is
\begin{eqnarray}
\omega _{2}^{bc} &=&c_{1}\partial _{m}\omega _{0}^{nl}\partial _{n}\omega
_{0}^{mk}\partial _{l}\partial _{k}\omega _{0}^{bc}+c_{2}\partial _{k}\omega
_{0}^{bm}\partial _{l}\omega _{0}^{cn}\partial _{n}\partial _{m}\omega
_{0}^{kl}  \label{43} \\
&&+c_{3}\partial _{n}\partial _{k}\omega _{0}^{bm}\partial _{m}\partial
_{l}\omega _{0}^{cn}\omega _{0}^{kl}~.  \notag
\end{eqnarray}%
After long calculations one finds that $c_{1}=-\frac{1}{12}$,
$c_2=0$ and $c_{3}=\frac{1}{6}$.
The same solution of the equation (\ref{42}) was obtained
in \cite{Penkava}. There are (at least) two cases when no
corrections to $\omega$ are needed. One is linear Poisson structures,
and indeed $\omega_2=0$, as follows from (\ref{43}). The other case
is deformations on a two-dimensional plane. In this latter case
$\omega_2$ calculated by formula (\ref{43}) is non-zero.
In fact, the consistency conditions are trivial in two dimensions,
and any function, $\omega_0$ or $\omega_2$, solves them. Since $\omega_0$
in two dimensions is not restricted by the Jacobi identities, (i.e., it is
completely general), the most natural choice in two dimensions is
$\omega_2=0$.

Now we are able to obtain the third order operator valued function
$\hat\omega_3^{ij}$ by collecting corresponding orders in 
$\alpha$ in the Weyl ordered expression 
$W(\omega_0^{ij}+\alpha^2\omega^{ij}_2)$. Explicitly, it reads
\begin{eqnarray}
&&\alpha^{-3}\hat\omega_{3}^{ij}=
\left( \left( \Gamma _{1}^{ijk}+\frac{1}{2}\Gamma _{0}^{jmn}\partial
_{m}\partial _{n}\omega _{0}^{ik}\right) \partial _{a}\partial _{b}\omega
_{0}^{ij}+\left( \Gamma _{0}^{ijlk}+\frac{2}{3}\Gamma _{0}^{ijm}\partial
_{m}\omega _{0}^{lk}+\right. \right.\nonumber  \\
&&\quad
\left. \frac{1}{3}\omega _{0}^{im}\partial _{m}\Gamma _{0}^{jlk}+\frac{1}{6%
}\omega _{0}^{im}\partial _{m}\omega _{0}^{jn}\partial _{n}\omega _{0}^{lk}+%
\frac{1}{6}\omega _{0}^{im}\omega _{0}^{jn}\partial _{m}\partial _{n}\omega
_{0}^{lk}\right) \partial _{a}\partial _{b}\partial _{l}\omega _{0}^{ij}
\nonumber \\
&&\quad
\left. +\omega _{0}^{lk}\partial _{l}\omega _{2}^{ij}+\omega
_{2}^{lk}\partial _{l}\omega _{0}^{ij}\right) \partial _{k}
\label{44} \\
&&\quad
+\left( \Gamma _{1}^{akl}\partial _{a}\omega _{0}^{ij}+\left( \frac{3}{2}%
\Gamma _{0}^{abkl}+\frac{1}{2}\omega _{0}^{am}\partial _{m}\Gamma
_{0}^{bkl}+\Gamma _{0}^{bmk}\partial _{m}\omega _{0}^{al}\right) \partial
_{a}\partial _{b}\omega _{0}^{ij}\right.\nonumber  \\
&&\quad
\left. +\left( \Gamma _{0}^{abk}\omega _{0}^{ml}+\frac{1}{2}\omega
_{0}^{bn}\partial _{n}\omega _{0}^{ak}\omega _{0}^{ml}\right) \partial
_{a}\partial _{b}\partial _{m}\omega _{0}^{ij}\right) \partial _{k}\partial
_{l}\nonumber \\
&&\quad
+\left( \Gamma _{0}^{bml}\omega _{0}^{ak}\partial _{a}\partial _{b}\omega
_{0}^{ij}+\frac{1}{6}\omega _{0}^{ak}\omega _{0}^{bl}\omega
_{0}^{nm}\partial _{a}\partial _{b}\partial _{n}\omega _{0}^{ij}\right. 
\nonumber \\
&&\quad
\left. +\Gamma _{0}^{aklm}\partial _{a}\omega _{0}^{ij}\right) \partial
_{k}\partial _{l}\partial _{m}~.\nonumber
\end{eqnarray}
From now on we have to distinguish between $\omega$ and $\omega_0$.
All $\Gamma$'s appearing in eq.\ (\ref{44}) and in the formulae
below are obtained by substituting $\omega=\omega_0$ in the expressions
from the previous sections.
\subsection{Fourth order star product}
In the previous section we have gained some experience in solving 
commutation relations, so that now we can move faster. The $\alpha^4$
terms in $\hat x^i$ read
\begin{equation}
\alpha^4 \gamma _{4}^{i}
=\alpha ^{4}\left( \Gamma _{2}^{ijk}\left( x\right) \partial
_{j}\partial _{k}+\Gamma _{1}^{ijkl}\left( x\right) \partial _{j}\partial
_{k}\partial _{l}+\Gamma _{0}^{ijklm}\left( x\right) \partial _{j}\partial
_{k}\partial _{l}\partial _{m}\right)\,, 
\label{45}
\end{equation}
where $\Gamma _{2}^{ijk}\left( x\right)$, $\Gamma _{1}^{ijkl}\left( x\right)$
and $\Gamma _{0}^{ijklm}\left( x\right)$ have to be determined. The formula
(\ref{bb}) for $n=3$ yields
\begin{eqnarray}
\alpha^{-4}G_{4}^{ijklm} &=&2\Gamma _{0}^{bml}
\omega _{0}^{ak}\partial _{a}\partial
_{b}\omega _{0}^{ij}+\frac{1}{3}\omega _{0}^{ak}\omega _{0}^{bl}\omega
_{0}^{nm}\partial _{a}\partial _{b}\partial _{n}\omega _{0}^{ij}+2\Gamma
_{0}^{aklm}\partial _{a}\omega _{0}^{ij} \\
&&+3\Gamma _{0}^{jlmn}\partial _{n}\omega _{0}^{ik}-3\Gamma
_{0}^{ilmn}\partial _{n}\omega _{0}^{jk}-2\Gamma _{0}^{imn}\partial
_{n}\Gamma _{0}^{jkl}+2\Gamma _{0}^{jmn}\partial _{n}\Gamma _{0}^{ikl} 
\notag \\
&&+\omega _{0}^{in}\partial _{n}\Gamma _{0}^{jklm}-\omega _{0}^{jn}\partial
_{n}\Gamma _{0}^{iklm}~,  \notag
\end{eqnarray}%
\begin{eqnarray}
\alpha^{-4}G_{4}^{ijkl} &=&\left( 3\Gamma _{0}^{abkl}+3\omega _{0}^{am}\partial
_{m}\Gamma _{0}^{bkl}+2\Gamma _{0}^{bmk}\partial _{m}\omega _{0}^{al}\right)
\partial _{a}\partial _{b}\omega _{0}^{ij} \\
&&+2\Gamma _{1}^{akl}\partial _{a}\omega _{0}^{ij}+\left( 2\Gamma
_{0}^{abk}\omega _{0}^{ml}+\omega _{0}^{bn}\partial _{n}\omega
_{0}^{ak}\omega _{0}^{ml}\right) \partial _{a}\partial _{b}\partial
_{m}\omega _{0}^{ij}  \notag \\
&&+3\Gamma _{0}^{jlmn}\partial _{m}\partial _{n}\omega _{0}^{ik}-3\Gamma
_{0}^{ilmn}\partial _{m}\partial _{n}\omega _{0}^{jk}-2\Gamma
_{1}^{ilm}\partial _{m}\omega _{0}^{jk}  \notag \\
&&+2\Gamma _{1}^{jlm}\partial _{m}\omega _{0}^{ik}+\omega _{0}^{in}\partial
_{n}\Gamma _{1}^{jkl}-\omega _{0}^{jn}\partial _{n}\Gamma _{1}^{ikl}  \notag
\\
&&-\Gamma _{0}^{imn}\partial _{m}\partial _{n}\Gamma _{0}^{jkl}+\Gamma
_{0}^{jmn}\partial _{m}\partial _{n}\Gamma _{0}^{ikl}~,  \notag
\end{eqnarray}%
\begin{eqnarray}
\alpha^{-4}G_{4}^{ijk} &=&\left( 2\Gamma _{1}^{ijk}+\Gamma _{0}^{jmn}\partial
_{m}\partial _{n}\omega _{0}^{ik}\right) \partial _{a}\partial _{b}\omega
_{0}^{ij}+\left( 2\Gamma _{0}^{ijlk}+\frac{4}{3}\Gamma _{0}^{ijm}\partial
_{m}\omega _{0}^{lk}\right.  \\
&&+\frac{2}{3}\omega _{0}^{im}\partial _{m}\Gamma _{0}^{jlk}+\frac{1}{3}%
\omega _{0}^{im}\partial _{m}\omega _{0}^{jn}\partial _{n}\omega _{0}^{lk} 
\notag \\
&&\left. +\frac{1}{3}\omega _{0}^{im}\omega _{0}^{jn}\partial _{m}\partial
_{n}\omega _{0}^{lk}\right) \partial _{a}\partial _{b}\partial _{l}\omega
_{0}^{ij}+2\omega _{0}^{lk}\partial _{l}\omega _{2}^{ij}+2\omega
_{2}^{lk}\partial _{l}\omega _{0}^{ij}  \notag \\
&&-\omega _{0}^{il}\partial _{l}\omega _{2}^{jk}+\omega _{0}^{jl}\partial
_{l}\omega _{2}^{ik}-\omega _{2}^{il}\partial _{l}\omega _{0}^{jk}+\omega
_{2}^{jl}\partial _{l}\omega _{0}^{ik}  \notag \\
&&-\Gamma _{1}^{ilm}\partial _{l}\partial _{m}\omega _{0}^{jk}+\Gamma
_{1}^{jlm}\partial _{l}\partial _{m}\omega _{0}^{ik}-\Gamma
_{0}^{ilmn}\partial _{l}\partial _{m}\partial _{n}\omega _{0}^{jk}  \notag \\
&&+\Gamma _{0}^{jlmn}\partial _{l}\partial _{m}\partial _{n}\omega _{0}^{ik}
\notag
\end{eqnarray}

By using Lemma \ref{L2}, one obtains
\begin{eqnarray}
&& \alpha ^{4}\Gamma _{0}^{ijklm}=\frac{1}{20}\left(
G_{4}^{ijklm}+G_{4}^{ikjlm}+G_{4}^{iljkm}+G_{4}^{imjkl}\right) ~,  \label{50}
\\ &&
\alpha ^{4}\Gamma _{1}^{ijkl}=\frac{1}{12}\left(
G_{4}^{ijkl}+G_{4}^{ikjl}+G_{4}^{ilkj}\right)  \label{51}\\
&&
\alpha ^{4}\Gamma _{2}^{ijk}=\frac{1}{6}\left(
G_{4}^{ijk}+G_{4}^{ikj}\right) ~.  \label{52}
\end{eqnarray}
Next, with the help of the formulae (\ref{d3}), (\ref{2}), (\ref{dec})
and (\ref{AB2})
we construct the fourth order star product. It can be conveniently
represented as 
\begin{equation}
f\star _{4}g=\alpha^4 \sum L_{m,n}\left( f,g\right) ~,  \label{53}
\end{equation}%
where each term in the sum has $m$ derivatives acting on $f$ and
$n$ derivatives acting on $g$. The forth order star product is
symmetric. Therefore,
\begin{equation}
L_{m,n}\left( f,g\right)=
L_{n,m}\left( g,f\right).\label{symL}
\end{equation}
All non-vanishing $L_{m,n}\left( f,g\right)$ with $m\le n$ are listed
below.
\begin{eqnarray*}
L_{1,2}\left( f,g\right)  &=&\partial _{i}f\Gamma _{2}^{ijk}\partial
_{j}\partial _{k}g~, \\
L_{1,3}\left( f,g\right)  &=&\partial _{i}f\Gamma _{1}^{ijkl}\partial
_{j}\partial _{k}\partial _{l}g~, \\
L_{1,4}\left( f,g\right)  &=&\partial _{i}f\Gamma _{0}^{ijklm}\partial
_{j}\partial _{k}\partial _{l}\partial _{m}g~, \\
L_{2,2}\left( f,g\right)  &=&\partial _{i}\partial _{m}f\left( \Gamma
_{1}^{ijl}\partial _{j}\omega _{0}^{mn}+\frac{3}{2}\Gamma
_{0}^{ijkl}\partial _{j}\partial _{k}\omega _{0}^{mn}\right.  \\
&&\left. +\omega _{0}^{ij}\partial _{j}\Gamma _{1}^{mnl}+\frac{1}{2}\Gamma
_{0}^{ijk}\partial _{j}\partial _{k}\Gamma _{0}^{mnl}+\frac{3}{2}\Gamma
_{1}^{imnl}\right) \partial _{l}\partial _{n}g~, \\
L_{2,3}\left( f,g\right)  &=&\partial _{i}\partial _{m}f\left( \frac{1}{2}%
\omega _{0}^{ij}\partial _{j}\Gamma _{0}^{mnkl}+\frac{1}{2}\Gamma
_{1}^{ilk}\omega _{0}^{mn}\right.  \\
&&\left. +\frac{3}{2}\Gamma _{0}^{ijkl}\partial _{j}\omega _{0}^{mn}+\Gamma
_{0}^{ijk}\partial _{j}\Gamma _{0}^{mnl}+2\Gamma _{0}^{ijklm}\right)
\partial _{k}\partial _{l}\partial _{n}g~, \\
L_{2,4}\left( f,g\right)  &=&\partial _{i}\partial _{m}f\left( \omega
_{0}^{ij}\Gamma _{0}^{mnkl}+\frac{1}{2}\Gamma _{0}^{ijkl}\Gamma
_{0}^{mnl}\right) \partial _{j}\partial _{k}\partial _{l}\partial _{n}g~, \\
L_{3,3}\left( f,g\right)  &=&\partial _{i}\partial _{j}\partial _{m}f\left( 
\frac{1}{2}\omega _{0}^{ia}\partial _{a}\omega _{0}^{jk}\Gamma _{0}^{mnl}+%
\frac{1}{2}\omega _{0}^{ia}\omega _{0}^{jk}\partial _{a}\Gamma
_{0}^{mnl}\right.  \\
&&+\left. \omega _{0}^{ik}\Gamma _{0}^{jal}\partial _{a}\omega _{0}^{mn}+%
\frac{2}{3}\Gamma _{0}^{mnl}\Gamma _{0}^{ijk}+\omega _{0}^{mn}\Gamma
_{0}^{ijkl}\right) \partial _{k}\partial _{l}\partial _{n}g~, \\
L_{3,4}\left( f,g\right)  &=&\frac{1}{2}\partial _{i}\partial _{j}\partial
_{m}f\omega _{0}^{ia}\omega _{0}^{jk}\Gamma _{0}^{mnl}\partial _{a}\partial
_{k}\partial _{l}\partial _{n}g~, \\
L_{4,4}\left( f,g\right)  &=&\frac{1}{24}\partial _{i}\partial _{j}\partial
_{m}\partial _{n}f\omega _{0}^{ik}\omega _{0}^{jl}\omega _{0}^{ma}\omega
_{0}^{nb}\partial _{k}\partial _{l}\partial _{a}\partial _{b}g~.
\end{eqnarray*}

At the fourth order, as in all even orders,
\begin{equation}
f\star_4 g - g\star_4 f =0, \label{f4g}
\end{equation}
so that the consistency condition is satisfied automatically, and $\omega$
must not be corrected. 

The fifth order star product follows simply by iterating the procedure
presented above. We do not present explicit expressions since they are
extremely lengthy. 
\section{Discussion and conclusions}
In this paper we developed an approach to the star products on $\mathbb{R}^N$
based on the Weyl symmetrically ordered operator products, which is a rather
natural language in physics. An important part of our scheme is a differential
representation of deformed coordinates $\hat x^j$. By using this representation
we were able to develop a rather effective iterative scheme where all
calculations of the higher orders of the star product go automatically,
besides solving the consistency condition at (some of the) odd orders
of the expansion. So far, we cannot give a general solution for (or a
method of solving of) this condition. This is an interesting problem
on its own right, especially given relations to the cohomology
theory, and we are going to address it in the future.

Explicit representations for NC coordinates are important in
NC quantum mechanics \cite{QM1,QM2} (see also \cite{QM3} for
more recent references). Applications to NC quantum mechanics
with a position-dependent noncommutativity was among main
motivations of our work.

If, for some reason, the consistency condition (\ref{conscond}) is satisfied
automatically at all orders (which is true, e.g., for deformations on 
$\mathbb{R}^2$ or for some polynomial algebras) our formulae provide
a star product at any finite order in $\alpha$, and, in principle, one
can even try to guess or derive a non-perturbative expression for the
star product. It would also be interesting to relate our iteration
formulae to the Ward identities of the path integral of Cattaneo and
Felder   \cite{Cattaneo:1999fm} with the help of the methods
\cite{Kummer:1996hy,Hirshfeld:1999xm} developed for calculation 
of such integrals.

Combining our approach with other approaches may also bring interesting
results. It is very natural to consider linear Poisson structures,
where simpler results are possible and a considerable progress has
already been made by the methods quite similar to the ones proposed
above \cite{Bordemann,Durov:2006iv,Meljanac:2006ui,Meljanac:2008ud,Chrys}.
For the reason we already mentioned, our formalism simplifies on
two dimensional Poisson manifolds, where a coherent state formalism
was used to construct star products \cite{Alexanian:2000uz}.
Another attractive possibility is to explore extensions of the
notion of quantum shift operator \cite{Chaichian:2005yp,Arai:2006zx}
to the case of a position dependent noncommutativity.
Finally, it is interesting and important to give our construction
a geometric flavor along the lines of \cite{Fedosov:1994zz} and
\cite{Batalin:1990yk,PT}.

\section*{Acknowledgements}
We are grateful to Alexander Pinzul for fruitful discussions and to
Giuseppe Dito and Daniel Sternheimer for correspondence. The work of
V.G.K.\ was supported by FAPESP. D.V.V.\ was supported in part by
FAPESP and CNPq.

\appendix
\section{Some useful formulae}\label{AppA}
In order to use the Duhamel expansion (\ref{dec}) one has to calculate
repeated commutators of $A=-ip_mx^m$ and $B=-ip_j(\hat x^j-x^j)$.
Denote $B_k=-ip_j\Gamma^{ji_1\dots i_k}\partial_{i_1}\dots \partial_{i_k}$.
Then
\begin{eqnarray}
&&[B_k,A]=k(-ip_j)(-ip_{i_1})
\Gamma^{ji_1i_2\dots i_k}\partial_{i_2}\dots \partial_{i_k},\quad k\ge 2,
\nonumber\\
&&[[B_k,A],A]=k(k-1)(-ip_j)(-ip_{i_1})(-ip_{i_2})
\Gamma^{ji_1i_2\dots i_k}\partial_{i_3}\dots \partial_{i_k},\quad k\ge 3,
\label{AB2}\\
&&[[[B_k,A],A],A]\nonumber\\
&&\quad =k(k-1)(k-2)(-ip_j)(-ip_{i_1})(-ip_{i_2})(-ip_{i_3})
\Gamma^{ji_1i_2\dots i_k}\partial_{i_4}\dots \partial_{i_k},\quad k\ge 4.
\nonumber
\end{eqnarray}
If $k$ does not satisfy the inequalities given above, 
the right hand side vanishes.
It is convenient to keep the momenta in the combinations $(-ip_m)$
since after taking the integral (\ref{2}) they are replaced by partial 
derivatives $\partial_m $ acting on $f$.

\end{document}